\documentclass{aip-cp}

\usepackage[numbers]{natbib}
\usepackage{rotating}
\usepackage{graphicx}
\newcommand{\beq}{\begin{equation}}
\newcommand{\eeq}{\end{equation}}
\newcommand{\bea}{\begin{eqnarray}}
\newcommand{\eea}{\end{eqnarray}}
\newcommand{\eps}{\varepsilon}

\begin{document}

% Title portion
\title{Particle-phonon coupling effects within theory of finite Fermi systems}

\author[aff1,aff2]{E.E. Saperstein \corref{cor1}}
%\eaddress[url]{http://www.aip.org}
%\eaddress{saper43\_7@mail.ru}
\author[aff1,aff3]{S.V. Tolokonnikov}
%\eaddress{tolkn@mail.ru}

\affil[aff1]{National Research Centre ``Kurchatov Institute'', 123182 Moscow, Russia.}
%). Note the use
%of superscript ``a)'' to indicate the author's e-mail address below. Use b), c), etc. to indicate
%e-mail addresses for more than 1 author.}
\affil[aff2]{National Research Nuclear University MEPhI, 115409 Moscow, Russia.} \affil[aff3]{Moscow
Institute of Physics and Technology, 123098 Dolgoprudny, Russia.} \corresp[cor1]{Corresponding author:
saper43\_7@mail.ru}

\maketitle

\begin{abstract}
Recent results of the study of the particle-phonon coupling (PC) effects in odd magic and semi-magic
nuclei within the self-consistent theory of finite Fermi systems are reviewed. In addition to the
usual pole diagrams, the non-pole ones are considered. Their contributions are often of a crucial
importance. PC corrections to the single-particle energies for $^{40}$Ca and $^{208}$Pb are presented.
The quadrupole moments of odd In and Sb isotopes, the odd-proton neighbors of even Sn isotopes, are
presented also with accounting for the PC corrections. At last, recently announced problem of
extremely high values charge radii of heavy Ca isotopes is solved in terms of a consistent
consideration of the PC effects. In all the cases, rather good description of the data is obtained.
\end{abstract}

% Head 1
\section{Introduction}

A consistent consideration of particle-phonon coupling (PC) effects within the self-consistent theory
of finite Fermi systems (TFFS) \cite{scTFFS} or any other self-consistent approach based on the use of
parameters fitted to experimental data is rather delicate problem. Indeed, the PC processes are
included, on average, to these phenomenological parameters. Therefore, the direct inclusion of all the
PC contributions to the predictions for nuclear characteristics should be inevitably accompanied with
a readjustment of the initial parameters.  To avoid such a time-consuming way, we try to consider only
the fluctuating part of the PC corrections which changes significantly from one nucleus to another
depending on the state $|\lambda\rangle$ of the odd nucleon and on the characteristics of the
$L$-phonon under consideration in the neighboring even-even nucleus.

A consistent method to describe the PC effects within the TFFS \cite{AB1} was developed by Khodel
\cite{Khod-76}. It was essentially based on the TFFS self-consistency relation \cite{Fay-Khod} between
the CM coordinate derivative of the mass operator $\Sigma(\eps,{\bf r}_1,{\bf r}_2)$, that of the
one-particle Green function $G(\eps,{\bf r}_1,{\bf r}_2)$, and the irreducible two-body interaction
block ${\cal U}(\eps,\eps';{\bf r}_1,{\bf r}_2,{\bf r}_3,{\bf r}_4)$. This relation is a consequence
of the spontaneous breaking of the translation symmetry in nuclei. In a simplified form, it reads \beq
\frac {\partial U} {\partial {\bf r}} = \int d{\bf r}' {\cal F} ({\bf r},{\bf r}') \frac {\partial
\rho} {\partial {\bf r}'}, \label{dUdR}\eeq  where $U({\bf r})$ is the mean-field nuclear potential,
$\rho({\bf r})$ is the nucleon density distribution, ${\cal F}$ being  the Landau--Migdal (LM)
interaction amplitude. The isotopic indices in (\ref{dUdR}) and below are for brevity omitted. In Ref.
\cite{Khod-76}, the low-lying natural parity excitations of even-even nuclei are interpreted as the
Goldstone mode arising due to the spontaneous breaking of the translation symmetry. The ``ghost''
dipole phonon, with zero excitation energy $\omega_1=0$, is the head of this branch, and the
corresponding vertex, as it follows from (\ref{dUdR}), is \beq g_1({\bf r})= \frac {\partial U}
{\partial r} Y_{1M}(\bf n)\label{g1},\eeq with obvious notation. According to \cite{Khod-76}, the
low-lying $L$-phonons are in many ways similar to the ghost one, their creation vertices possessing a
dominant surface behavior, \beq g_L({\bf r})= \left( \alpha_L \frac {\partial U} {\partial r} + \chi_L
(r)\right) Y_{LM}(\bf n)\label{gL},\eeq where $\alpha_L$ ia a coefficient showing the amplitude of the
surface $L$-vibration, typical values being $\simeq 0.3$ fm, and $\chi_L (r)$ is an in-volume addendum
which is rather small. Fig. 1 demonstrate the latter for the first $3^-$ state in the magic nucleus
$^{208}$Pb. Putting $\chi_L{=}0$, one obtains the prescription of the Collective Model by Bohr and
Mottelson (BM) \cite{BM2}, $\chi_L (r)$ being a quantum correction to this model. Fig. 1 represents
the result of solving the TFFS equation for $g_L$ with the use of the basis and LM amplitude generated
by the Fayans energy density functional (EDF) DF3-a \cite{DF3-a}, which is a version of the initial
Fayans EDF DF3 \cite{Fay}.

\begin{figure}[h]
  \centerline{\includegraphics[width=250pt]{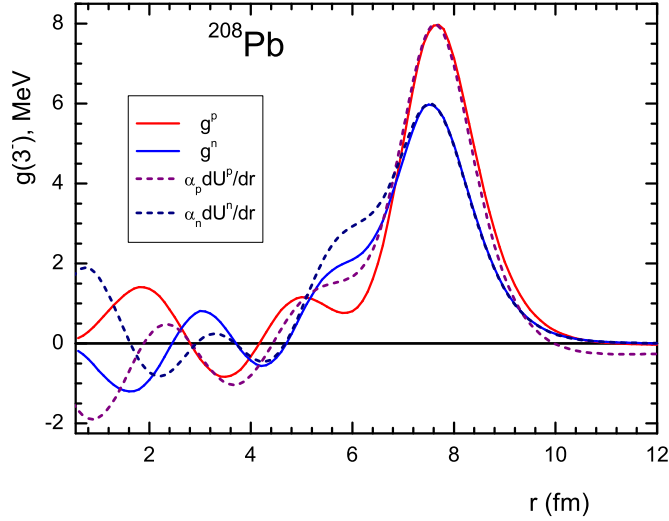}}
  \caption{$g_L$ vertex for the $3^-_1$ state in $^{208}$Pb, $\alpha_L^p
=0.334$ fm, $\alpha_L^n=0.322$ fm.}
\end{figure}

Usually, the perturbation theory in $g_L^2$ is used when PC corrections to observables are considered
\cite{AB1,Ring-Sch}, and the simplest, pole diagram is considered only. A principal feature of the
method developed in \cite{Khod-76} is inclusion into consideration of all non-pole diagrams of the
same $g_L^2$ order. Their sum does not depend on the energy $\eps$. However, at small $L$-phonon
excitation energy $\omega_L$, its contribution behaves as $1/\omega_L$, just as that of the pole
diagram. These two contributions possess opposite signs. As the result, their sum is regular at
$\omega_L\to 0$, in contrast to the case when the pole diagram is considered only. The method to
consider PC corrections in \cite{Khod-76} was developed in such a way that, being applied to the ghost
$1^-$ phonon, it results, as a rule, in the zero contributions. In the case of the total binding
energy $E_0$ or the single-particle (SP) energies $\eps_{\lambda}$, it is not zero, but very small
($\sim 1/A$, $A$ being the nuclear mass number), representing the recoil effect due to the CM motion.

In this review, we present briefly the results of \cite{Levels} for PC corrections to the SP energies
in magic nuclei and those of \cite{PC-Q} for PC corrected quadrupole moments of odd In and Sb
isotopes, the odd-proton neighbors of even Sn isotopes. At last, the results of \cite{Ca-Rch} are
given for the PC corrections to the charge radii of heavy Ca isotopes, where a ``puzzle'' announced
recently in \cite{exp-Ca-Rch} is resolved.

\section{Single-particle energies}
To find the SP energies with account for the PC effects, one must solve the following equation : \beq
\left(\eps-H_0 -\delta \Sigma^{\rm PC}(\eps) \right) \phi =0, \label{sp-eq}\eeq where $H_0$ is the
quasiparticle Hamiltonian with the spectrum $\eps_{\lambda}^{(0)}$ and $\delta \Sigma^{\rm PC}$ is the
PC correction to the quasiparticle mass operator.

Till now, all the self-consistent considerations of the PC effects within the Skyrme--Hartree--Fock
(SHF) method we know were limited to the problem of the SP energies  in magic nuclei
\cite{Bort,Dobaczewski,Baldo-PC}. Mention also a more recent consideration of this problem within the
relativistic mean-field theory \cite{Litv-Ring}. In all these works, the $g_L^2$ perturbation theory
for the phonon corrections to the mass operator $\Sigma$, $g_L$ being the vertex of creation of the
$L$-phonon, and the pole diagram was considered only, the first one in Fig. 2. In \cite{Levels}, the
problem was considered on the base of the self-consistent TFFS \cite{scTFFS} for seven magic nuclei
including the ``new'' mags \cite{SPE-exp}. The non-pole diagrams were included into consideration, see
the second term in Fig. 2. Note that often, see e.g. \cite{tad-KS},  a sum of the non-pole diagrams is
named the ``phonon tadpole'', as an analogue of the tadpole diagrams in the field theory
\cite{tad-Wein}.

In magic nuclei we deal, the vertex $g_L=\delta_L \Sigma$ obeys the equation \cite{AB1} \beq {
g_L}(\omega)={{\cal F}} {A}(\omega) { g_L}(\omega), \label{eqg_L} \eeq
 where $ A(\omega)=\int
G \left(\eps + \omega/ 2 \right) G \left(\eps - \omega/ 2 \right)d \eps/(2 \pi i)$ is the
particle-hole propagator. In obvious symbolic notation, the pole diagram corresponds to
$\delta\Sigma^{\rm pole}=(g_L,D_L g_L)$, where $D_L(\omega)$ is the $L$-phonon $D$-function.
Explicitly one obtains \beq \delta\Sigma^{\rm pole}_{\lambda\lambda}(\epsilon)=\sum_{\lambda_1\,M}
|\langle\lambda_1|g_{LM}|\lambda\rangle|^2 \left(\frac{n_{\lambda_1}}{\eps+\omega_L-
\eps_{\lambda_1}^{(0)}}+\frac{1-n_{\lambda_1}}{\eps-\omega_L -\eps_{\lambda_1}^{(0)}}\right),
\label{dSig2} \eeq where   $n_{\lambda}=(0,1)$ stands for the occupation numbers.

The second, non-pole, term in Fig. 2  is \beq \delta\Sigma^{\rm non-p}=\int \frac {d\omega} {2\pi i}
\delta_L {g_L} D_L(\omega),\label{tad} \eeq where $\delta_L {g_L}$  can be found \cite{scTFFS,Khod-78}
by variation of Eq. (\ref{eqg_L}) in the field of the $L$-phonon: \beq  \delta_L {g_L}=\delta_L {\cal
F} A(\omega_L) g_L + {\cal F} \delta_L A(\omega_L) g_L + {\cal F} A(\omega_L)\delta_L g_L. \label{dgL}
\eeq

As it was discussed in the Introduction, all the $L$-phonons we consider are of surface nature, the
surface peak dominating in their creation amplitude (\ref{gL}), see e.g. Fig. 1.
 If we neglect the the in-volume term $\chi_L$, a simple form of the non-pole term of the PC
 correction to the
mass operator $\Sigma$ can readily be obtained: \beq \delta\Sigma^{\rm non-p}_L(r) = \frac {\alpha_L
^2} 2 \frac {2L+1} 3 \triangle U(r). \label{tad}\eeq Its contribution to the SP energies is, as a
rule, comparable in the absolute value with that of the usual pole diagram, being usually of the
opposite sign. In the result, neglect of the tadpole PC term results often in an overestimate of the
PC correction to the SP energies. It is demonstrated in Table 1 for $^{40}$Ca and Table 2 for
$^{208}$Pb, which are based on the results of \cite{Levels}. All calculations were carried out with
the Fayans EDF DF3-a. The perturbation theory (PT) in $\delta \Sigma$ with respect to $H_0$ was used
for solving Eq. (\ref{sp-eq}). Smallness of the differences of $(1{-}Z^{\rm PC}_{\lambda})$ is a
criterium of validity for PT, where \beq Z_{\lambda}^{\rm PC} =\left({1- \left(\frac {\partial}
{\partial \eps}
 \delta \Sigma^{\rm PC}(\eps) \right)_{\eps=\eps_{\lambda}^{(0)}}}\right)^{-1}. \label{Z-fac}\eeq
In all magic nuclei considered in \cite{Levels}, typical values of $(1{-}Z^{\rm PC}_{\lambda})$ is
$0.1\div 0.2$, hence the PT is valid with high accuracy.

\begin{figure}
  \centerline{\includegraphics[width=250pt]{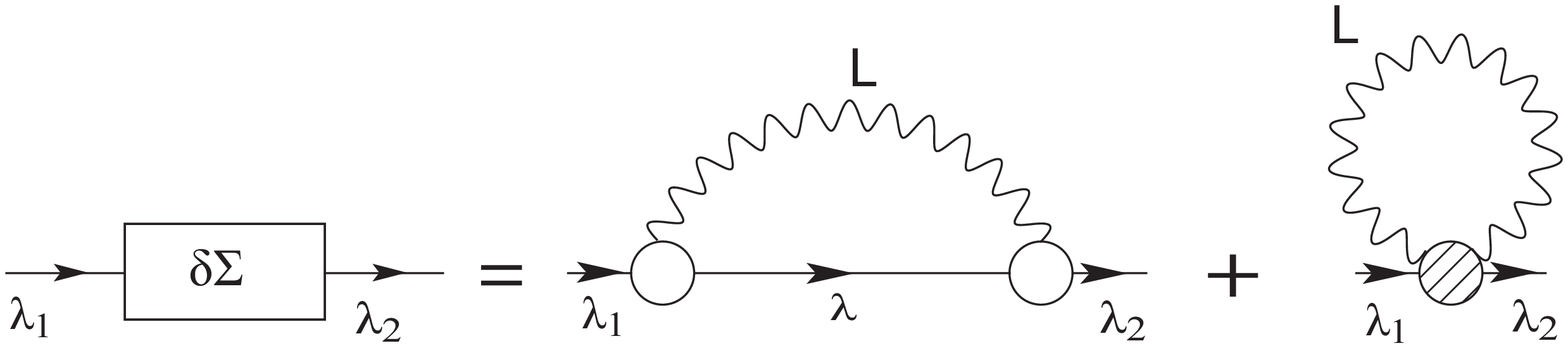}}
  \caption{$g_L^2$ phonon corrections to the mass operator. The dashed blob denotes the sum of non-pole
  (``phonon tadpole'') diagrams.}
\end{figure}

\begin{table}
\caption{Pole and non-pole contributions to PC corrections from $3^-$-states to SP energies (MeV) in
$^{40}$Ca, $\delta \eps_{\lambda}(1^-)$ is the PC correction due to the spurious $1^-$ phonon, i.e.
the CM recoil effect. $\eps_{\lambda}^{(0)}$ and $\eps_{\lambda}$ are the SP energies without and with
PC corrections, correspondingly.}
%\begin{center}
\begin{tabular}{cccccccc}
\noalign{\smallskip}\hline\noalign{\smallskip}

$\lambda$ &$\delta \eps^{\rm pole}_{\lambda}$ & $\delta \eps^{\rm non-p}_{\lambda}$ & $\delta
\eps_{\lambda}(3^-)$
& $\delta \eps_{\lambda}(1^-)$& $\eps_{\lambda}^{(0)}$ & $\eps_{\lambda}$
& $\eps_{\lambda}^{\rm exp}$\cite{SPE-exp}\\
\noalign{\smallskip}\hline\noalign{\smallskip} \noalign{\smallskip}
         &        &   & neutr.     &    &     & &\\
\noalign{\smallskip}
$1f_{5/2}$ &-0.395 &  0.592 &  0.197 &  0.321 &-2.124  & -1.634 & -2.65 \\
$2p_{1/2}$ &-0.805 &  0.305 & -0.500 &  0.133 &-3.729  & -4.072 & -4.42\\
$2p_{3/2}$ &-0.833 &  0.383 & -0.450 &  0.130 &-5.609  & -5.902 & -6.42\\
$1f_{7/2}$ &-0.142  &  0.733 & 0.591 &  0.173 &-9.593  & -8.870 & -8.36\\
$1d_{3/2}$ &-0.426 &  0.697 &  0.271 &  0.267 &-14.257 & -13.738& -15.64 \\
$2s_{1/2}$ &-0.932 &  0.493 & -0.439 &  0.184 &-15.780 & -16.017& -18.11\\
$1d_{5/2}$ &-0.253 &  0.731 &  0.478 &  0.224 &-19.985 & -19.305&  -21.27\\

\noalign{\smallskip}\hline\noalign{\smallskip}
         &  &      &   prot.     &   &   &   & \\
\noalign{\smallskip}
$1f_{5/2}$ &-0.240 &  0.470 &  0.230 &  0.300 & 4.359  & 4.869 & 4.60 \\
$2p_{1/2}$ &-0.584 &  0.152 & -0.432 &  0.062 & 2.456  & 2.104& 2.38 \\
$2p_{3/2}$ &-0.224 &  0.251 &  0.027 &  0.091 & 0.936  &  1.050& 0.63 \\
$1f_{7/2}$ & 0.100  & 0.677  &-2.678 & -0.198 &  0.777& -2.122& -1.09 \\
$1d_{3/2}$ &-0.370 &  0.659 &  0.289 & 0.262  &-7.264 & -6.733& -8.33 \\
$2s_{1/2}$ &-0.886 &  0.429 & -0.457 & 0.170  &-8.663 & -8.931& -10.85 \\
$1d_{5/2}$ &-0.234 &  0.699 &  0.466 & 0.216  &-12.856& -12.196&  -13.73 \\

\hline

\end{tabular}\label{tab:Ca-tad}
\end{table}

\begin{table}
\caption{Pole and non-pole contributions to PC corrections from $3^-$-states to SP energies (MeV) in
$^{208}$Pb. The quantity $\delta \eps_{\lambda}^{\rm tot}$ involves the PC corrections from $3^-_1$
and 8 other phonons, two $5^-$-states and 6 positive parity states (two $2^+$, two $4^+$, and two
$6^+$), the recoil correction is also taken into account, but it is very small for Pb and other  heavy
nuclei. }

\begin{tabular}{cccccccc}
\noalign{\smallskip}\hline\noalign{\smallskip}

$\lambda$ &  $\delta \eps^{\rm pole}_{\lambda}(3^-)$ & $\delta \eps^{\rm non-p}_{\lambda}(3^-)$ &
$\delta \eps_{\lambda}(3^-)$ & $\delta \eps_{\lambda}^{\rm tot}$
& $\eps_{\lambda}^{(0)}$ & $\eps_{\lambda}$ & $\eps_{\lambda}^{\rm exp}$\cite{SPE-exp}\\
\noalign{\smallskip}\hline\noalign{\smallskip}
         &        &   & neutr.     &    &     & &\\
\noalign{\smallskip}
$3d_{3/2}$ & -0.150 &  0.012 & -0.137 &-0.462 &-0.709  &  -1.171 & -1.40 \\
$2g_{7/2}$ & -0.142 &  0.061 & -0.081 &-0.335 &-1.091  &  -1.426 & -1.45 \\
$4s_{1/2}$ & -0.134 &  0.016 & -0.118 &-0.403 &-1.080  &   -1.483 & -1.90\\
$3d_{5/2}$ & -0.147 &  0.023 & -0.124 &-0.424 &-1.599  &   -2.023 & -2.37 \\
$1j_{15/2}$& -0.708 &  0.204 & -0.504 &-0.316 &-2.167  &  -2.483 & -2.51 \\
$1i_{11/2}$& -0.058 &  0.198 &  0.140 & 0.184 &-2.511  &  -2.327 & -3.16 \\
$2g_{9/2}$ & -0.244 &  0.076 & -0.167 &-0.250 &-3.674  &   -3.924 & -3.94 \\
$3p_{1/2}$ & -0.220 &  0.053 & -0.167 &-0.043 &-7.506  &  -7.549 & -7.37\\
$2f_{5/2}$ & -0.186 &  0.094 & -0.092 & 0.114 &-8.430  &  -8.316 & -7.94 \\
$3p_{3/2}$ & -0.205 &  0.056 & -0.149 & 0.025 &-8.363  &  -8.338 & -8.27\\
$1i_{13/2}$&  0.057 &  0.211 &  0.269 & 0.506 &-9.411  &  -8.905 & -9.00\\
$2f_{7/2}$ &  0.724 &  0.091 &  0.815 & 0.649 &-10.708 &  -10.059 & -9.71\\
$1h_{9/2}$ & -0.014 &  0.197 &  0.184 & 0.474 &-11.009  &  -10.535 & -10.78\\

\noalign{\smallskip}\hline\noalign{\smallskip} \noalign{\smallskip}
          &        &   & prot.           &    &     & &  \\
\noalign{\smallskip}
$3p_{1/2}$ & -0.375 &  0.153 & -0.222 & 0.000  &0.484  &    0.484 & -0.17\\
$3p_{3/2}$ & -0.371 &  0.152 & -0.219 &-0.561  &-0.249 &   -0.810 & -0.68\\
$2f_{5/2}$ & -0.278 &  0.168 & -0.110 &-0.361  &-0.964 &   -1.325 & -0.98\\
$1i_{13/2}$& -0.534 &  0.266 & -0.268 &-0.152  &-2.082 &   -2.234 & -2.19 \\
$2f_{7/2}$ & -0.409 &  0.168 & -0.240 &-0.291  &-3.007 &   -3.298 & -2.90\\
$1h_{9/2}$ & -0.054 &  0.222 &  0.168 & 0.273  &-4.232 &    -3.959 & -3.80 \\
$3s_{1/2}$ & -0.310 &  0.143 & -0.167 &-0.022  &-7.611 &   -7.633 & -8.01 \\
$2d_{3/2}$ & -0.241 &  0.146 & -0.095 & 0.060  &-8.283 &   -8.223 & -8.36 \\
$1h_{11/2}$& -0.017 &  0.246 &  0.229 & 0.411  &-8.810 &    -8.399 & -9.36 \\
$2d_{5/2}$ &  0.435 &  0.147 &  0.582 & 0.548  &-9.782 &    -9.234 & -9.70 \\
$1g_{7/2}$ & -0.271 &  0.197 & -0.074 & 0.122  &-11.735 &   -11.613 & -11.49 \\

\noalign{\smallskip}\hline\noalign{\smallskip}
\end{tabular}
\label{tab:Pb-tad}
\end{table}

Let us begin from $^{40}$Ca, Table 1. We see that always the non-pole correction $\delta \eps^{\rm
non-p}_{\lambda}$ is positive, whereas, with one exception of the proton state $1f_{7/2}$, the pole
one, $\delta \eps^{\rm pole}_{\lambda}$, is negative. There are cases, e.g. the neutron state
$2p_{3/2}$ or the proton one $2s_{1/2}$,  when the absolute value of  $\delta \eps^{\rm
pole}_{\lambda}$ is bigger than that of $\delta \eps^{\rm non-p}_{\lambda}$, and the total correction
is approximately two times less than the pole one alone. There are cases, e.g. the neutron $1f_{5/2}$
state or proton $1d_{3/2}$ one when the absolute value of $\delta \eps^{\rm non-p}_{\lambda}$ is
bigger and the total PC correction possesses the opposite sign, compared to the pole one. Note that
there is no cases when the non-pole correction is negligible.

Let us turn to the $^{208}$Pb nucleus, Table 2. We see that, for the $3_1^-$ state, all conclusions
about the role of the non-pole term of the PC correction, made above for $^{40}$Ca, remain valid. Now
8 phonons are taken into account, in addition to $3_1^-$, and their common contribution is
approximately equal to the one of the $3_1^-$ alone. For this nucleus, account for the PC corrections
to the SP energies makes agreement with experimental data \cite{SPE-exp} essentially better. Indeed,
the average deviation from the experimental values $\langle\delta \eps_{\lambda}\rangle_{\rm
rms}{=}0.51\;$MeV for the DF3-a EDF without PC corrections and 0.34 MeV with PC corrections. Note,
that, at the mean field level, DF3-a surpasses popular Skyrme EDFs, e.g. for the HFB-17 EDF, for the
$^{208}$Pb nucleus one gets $\langle\delta \eps_{\lambda}\rangle_{\rm rms}{=}1.15\;$MeV, as it was
found in \cite{Levels}.

\section{Electro-magnetic moments}

In the last two decades, a lot of new data on the electromagnetic moments of ground and isomeric
states of odd spherical nuclei have appeared due to the intensive work of several modern radioactive
ion beam facilities. As a result, the bulk of the data on nuclear static moments has become very
extensive and comprehensive \cite{Stone}, creating a challenge to nuclear theory. Because of technical
problems, the analysis of these data within the many-particle Shell Model \cite{SM1,SM2} was limited
with light and intermediate mass nuclei, $A{<}90$. For the whole nuclear map, this challenge was
partially responded within the self-consistent theory TFFS for magnetic moments \cite{mu-YAF,mu-EPJA}
and quadrupole moments \cite{BE2,Q-EPJA}. The consideration was carried out at the mean-field level
with the use of the Fayans EDF DF3 \cite{Fay} for magnetic moments and DF3-a \cite{DF3-a}, for
quadrupole ones. With a good overall description of the data, there are cases of noticeable deviations
which can be naturally attributed to the PC effects.

The model for describing the PC corrections to electromagnetic moments was developed first in
\cite{EPL-2013}. The main idea was to separate and explicitly consider such PC diagrams that behave in
a non-regular way, depending significantly on the nucleus under consideration and the SP state
$|\lambda\rangle$ of the odd nucleon. The rest (and the major part) of the PC corrections is supposed
to be regular and included in the initial values of the TFFS parameters, in particular, those of the
LM amplitude. Firstly, this model was applied to the magnetic moments \cite{EPL-2013,PC-mu-YAF} and
recently, to the quadrupole ones \cite{PC-Q}. The PC diagrams included into the model are displayed in
figures 3, 4 for the case of the qudrupole moments, the external field being of the $E2$ type. The
model is developed for semi-magic nuclei, which contain a superfluid subsystem and a normal one, and
besides the odd nucleon belongs to the normal subsystem. It simplifies the problem drastically. In
addition, a non-regular behavior of the PC corrections is typical namely for the normal subsystem  of
a semi-magic nucleus.

\begin{figure}[h!]
 \hspace{-10mm}
  \includegraphics[width=150pt]{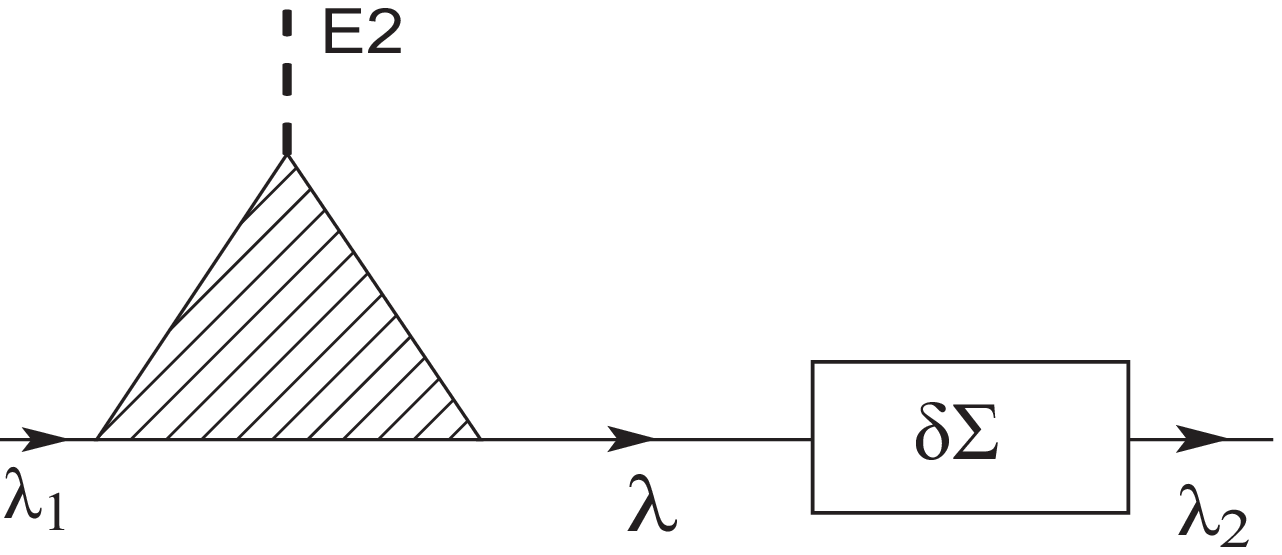}
  \hspace{10mm}
    \includegraphics[width=100pt]{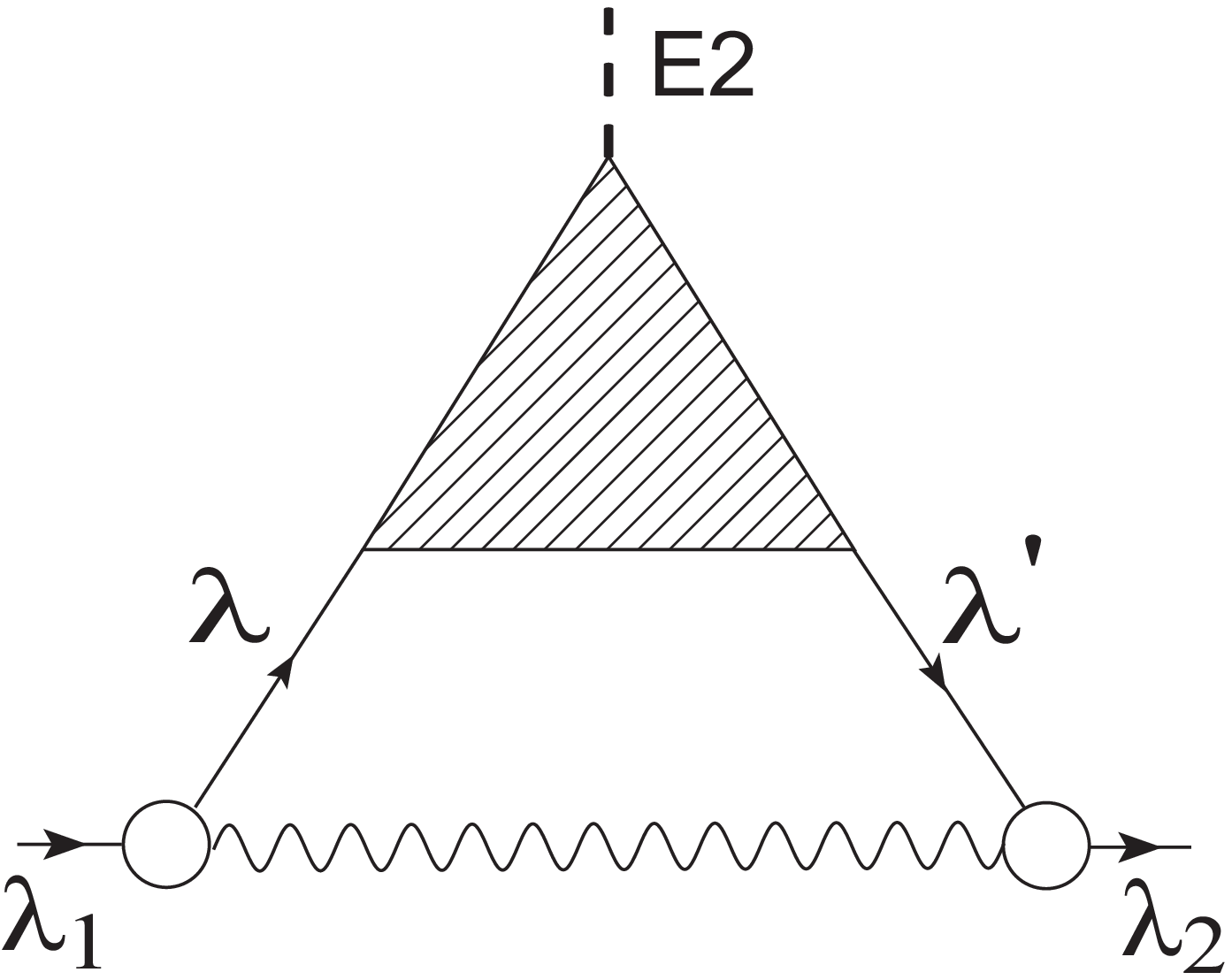}
  \caption{Diagrams for two PC corrections to the quadrupole moment of an odd nucleus:
  the ``end correction'' (left) and the one due to the induced interaction (right).}
\end{figure}

The left part of Fig. 3 illustrates so-called ``end correction''. Of course, there is a symmetric
diagram with similar correction to the left end. The main end correction occurs in  the diagonal case,
$\lambda{=}\lambda_2$, when the corresponding expression possesses a pole. There is the well-known
recipe to solve this problem  \cite{Levels}  by the renormalizing the end: $|\lambda_2\rangle\to
\sqrt{Z_{\lambda_2}}|\lambda_2\rangle$. In similar way, one gets $|\lambda_1\rangle\to
\sqrt{Z_{\lambda_1}}|\lambda_1\rangle$.

The rest of these sums with non-diagonal terms $\lambda_1 \neq \lambda$ can be calculated directly and
is rather small. However, we retain it for completeness, and represent the ``end correction'' to the
quadrupole moment value as the sum: \beq  \delta Q_{\lambda\lambda}^{\rm end} = \delta
Q_{\lambda\lambda}^Z + (\delta Q_{\lambda\lambda}^{\rm end})', \label{end3}  \eeq where \beq
\delta{Q}^Z_{\lambda\lambda} = \left(Z_{\lambda}-1\right) Q_{\lambda\lambda}. \label{endZ} \eeq Eqs.
(\ref{end3}), (\ref{endZ}) correspond
 to partial summation of the diagrams of Fig. 2, and, hence, contain  higher order  terms in $g_L^2$.
 To be consistent up to the order $g_L^2$, the $Z$-factors in
Eqs. (\ref{Z-fac}) and (\ref{endZ}) should be expanded in terms of $\partial
\Sigma_{\lambda\lambda}(\eps) /
\partial \eps$, $Z_{\lambda}^{\rm PT}=1+ \partial \delta \Sigma_{\lambda\lambda}(\eps) /
\partial \eps$, with the result
\beq \left(\delta{Q}^Z_{\lambda\lambda}\right)_{\rm PT} {=}  \left. \frac  {\partial \delta
\Sigma_{\lambda\lambda}(\eps)}{\partial \eps}\right|_{\eps=\eps_{\lambda}}  Q_{\lambda\lambda}.
\label{Z_PT} \eeq The energy derivative of the mass operator (\ref{dSig2}) can be readily found.

The PC correction due to the induced interaction is displayed in the right part of Fig. 3. As we shall
see, each of the two above PC corrections is rather big, but, being of opposite signs, they cancel
each other to large extent. Therefore, the complete self-consistency of the calculation scheme is very
important in this point.

\begin{figure}[h!]
 \hspace{-30mm}
  \includegraphics[width=120pt]{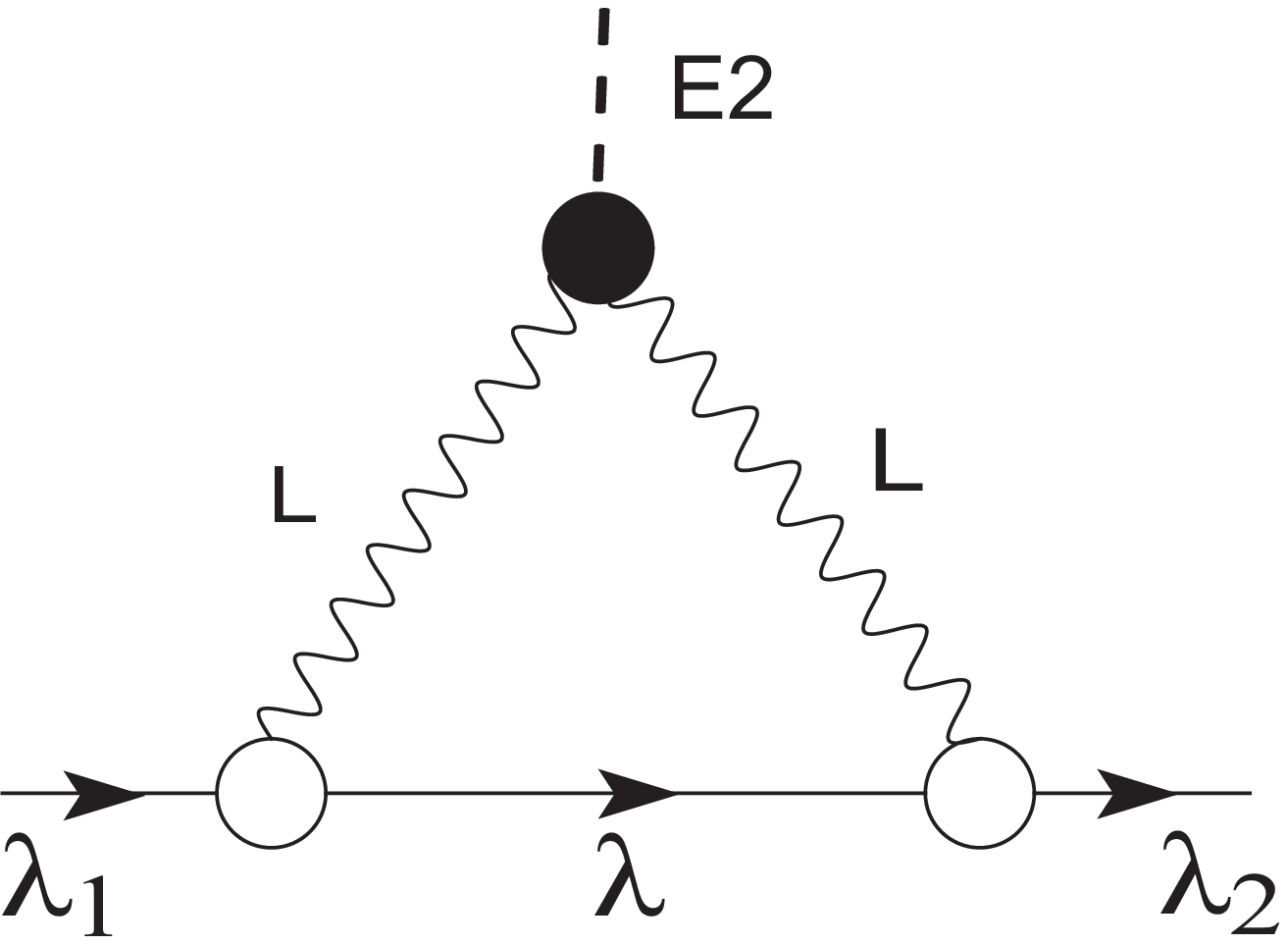}
  \hspace{30mm}
  \includegraphics[width=40pt]{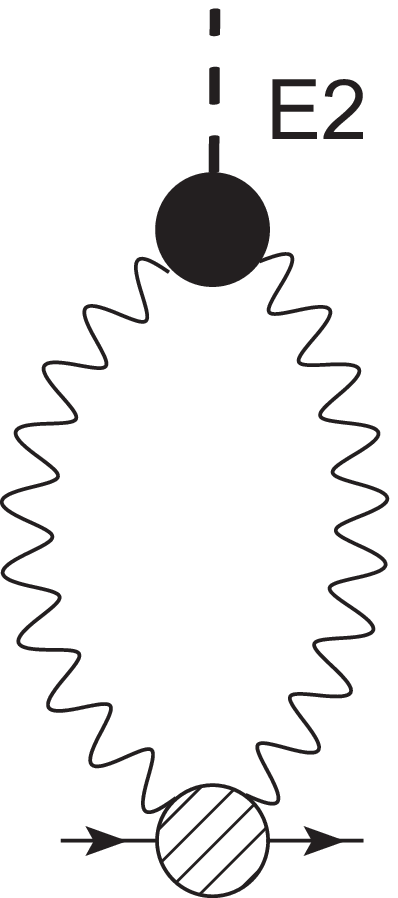}
  \caption{Diagrams for the PC correction due to the quadrupole moment of the $L$-phonon: the triangle
(GDD) diagram (left) and the non-pole one (right).}
\end{figure}

Let us consider the PC correction, the left part of Fig. 4, due to the quadrupole moment of the
$L$-phonon. We present briefly the results of \cite{PC-Q}, were the odd isotopes of In and Sb which
are the odd-proton neighbors of even Sn isotopes and the PC corrections due to the $2^+_1$ phonons
were considered only, $L{=}2$. After separating the angular variables for the ``triangle'' (GDD), with
the use of the short notation $\lambda_1 \to 1$, we obtain \beq \delta Q_{\lambda\lambda}^{GDD}
=(-1)^{j-m} \!\left(\!\begin{array}{ccc}\! j&\!\!2\!\!& j\!\\\!-m&\!\! M\!\!&
m\!\end{array}\right)\langle \nu \!\parallel \delta Q^{GDD}
\parallel \nu \rangle,\label{GDD1}\eeq
with the reduced matrix element \beq \langle 0 \parallel \delta Q^{GDD}
\parallel 0\rangle  {=} \sum_1 (-1)^{L+j_0+j_1} Q_L^{\rm ph}
\sqrt{ \frac {L(L+1)(2L+1)}{4\pi}} \left\{\!\begin{array}{ccc}j_0&2&j_0\\L&j_1&L\end{array}\right\}
\langle 1\!\parallel g_L
\parallel \!0\rangle  \langle 0\! \parallel \tilde g_L \parallel\! 1\rangle
\left(I_1^{(1)}
 (\omega_L)+I_1^{(2)}(\omega_L) \right),
  \qquad\qquad \label{GDD2}\eeq

\beq I_1^{(1)}(\omega_L) = \frac {1-n_1}{(\eps_0-\eps_1-\omega_L)^2} +  \frac
{n_1}{(\eps_0-\eps_1+\omega_L)^2}, \label{GDD3}\eeq \beq I_1^{(2)}(\omega_L) {=} -\frac 1 {\omega_L}
\left( \frac {n_1}{\eps_0{-}\eps_1{+}\omega_L} + \frac {1-n_1} {\eps_0{-}\eps_1{-}\omega_L} \right).
\label{GDD4}\eeq
 The second integral (\ref{GDD4}) reveals a dangerous behavior at $\omega_L \to 0$.
The non-pole diagram, the right part of Fig. 4, possess a similar singularity
\cite{EPL-2013,PC-mu-YAF}. Let us denote the corresponding terms of (\ref{GDD2}) with (\ref{GDD3}) and
(\ref{GDD4}) as $\delta\overline{} V^{(1),(2)}_{GDD}$. An ansatz was proposed in the model under
discussion how to deal with these two dangerious terms of $\delta Q_{GDD}$. It was supposed that the
term $\delta Q^{(2)}_{GDD}$ and the non-pole one $\delta Q_{\rm non-p}$ cancel each other. Such
cancelation does take place for the ``fictitious'' external fields $V_0={\bf j}$ and $V_0=1$ providing
the conservation of the total momentum of the system and the total particle number, correspondingly.
In addition, this is true in the case of the spurious $1^{-}$ phonon. In the result, the total $g_L^2$
PC correction  to the effective quadrupole field becomes equal to \beq \delta Q= \delta Q^Z_{\rm PB} +
\delta Q_{GGD} + \delta Q^{(1)}_{GDD}+ \delta Q'_{\rm end}. \label{PC-sum} \eeq

\begin{table}
\caption{Different PC corrections to the quadrupole moments of several odd In and Sb nuclei. $Q$ is
the quadrupole moment without PC corrections \cite{Q-EPJA}. Other notation is explained in the text.
All values, except $Z$, are in b.}
\begin{tabular}{|l| c| c| c| c| c| c| c| c| c|}
\noalign{\smallskip}\hline\noalign{\smallskip}    nucl.  &$\lambda$ & $Q$ & $Z$ &$\delta Q^Z_{\rm
ptb}$& $\delta Q_{GGD}$ & $\delta Q_{GDD}$ &
$\delta Q_{\rm end}'$ & $\delta Q_{\rm PB} $  & $\delta Q_{\rm ph} $\\
\noalign{\smallskip}\hline\noalign{\smallskip}

$^{109}$In & $1g_{9/2}$ & +1.113 &0.573 &-0.826 &0.487 &0.128 & 0.023 &-0.188 &-0.108\\

$^{111}$In & $1g_{9/2}$ & +1.165 &0.488 &-1.220&0.722 &0.163  & 0.034 &-0.301 &-0.147\\

$^{113}$In & $1g_{9/2}$ & +1.117 &0.576 &-0.820 &0.484 &0.071 & 0.025 &-0.240 &-0.138\\

$^{117}$Sb & $2d_{5/2}$&  -0.817  &0.582 &0.588  &-0.229 &-0.009    &0.050 &0.399 &0.232 \\

$^{119}$Sb & $2d_{5/2}$&  -0.763  &0.602 &0.504  &-0.198 &-0.001 &0.048   &0.353 &0.213 \\

$^{121}$Sb & $2d_{5/2}$&  -0.721  &0.591 &0.497  &-0.196 &0.003  &0.052   &0.355 &0.210 \\

\noalign{\smallskip}\hline\noalign{\smallskip}
\end{tabular}
\label{tab:delQ_ph}
\end{table}

As an example, separate terms of Eq. (\ref{PC-sum}) for the PC correction to the $Q$ value are given
in Table 3 for several nuclei under consideration. The complete table may be found in \cite{PC-Q}. We
see that two main corrections are those due to the $Z$-factor (column 5) and due to the induced
interaction (the term $\delta Q_{GGD}$, column 6). They always possess different signs, the sum being
significantly less in the absolute value than each of them. Therefore two other ``small'' corrections
are sometimes also important.

The results for PC-corrected values of the quadrupole moments of the In and Sb isotopes under
consideration are presented in Table 4 and Fig. 5. The mean-field predictions of \cite{Q-EPJA} are
given for comparison. We see that the PC corrections to quadrupole moments taken into account make
agreement with experiment better in most cases.  The rms value $\langle\delta \tilde{Q}\rangle_{\rm
rms}=0.15\;$b follows from the last column of Table 3. The corresponding value without PC corrections
is significantly bigger, $\langle\delta {Q}\rangle_{\rm rms}=0.27\;$b.

\begin{table}
 \caption{Quadrupole moments $Q\;$(b) of odd In and Sb isotopes. Experimental data are
taken from the review article \cite{Stone}. For the $^{115}$In isotope, the first value corresponds to
the original experiment of \cite{exp-In115-1}, the second one, to \cite{exp-In115-2}. Similarly, for
the $^{121}$Sb isotope, the first value corresponds to \cite{exp-Sb121-1}, the second one, to
\cite{exp-Sb121-2}. $Q_0$ is the prediction of the single-particle model. The theoretical values are
$Q_{\rm th}$ and $\tilde{Q}_{\rm th}$ without and with PC corrections, correspondingly. The
differences $\delta Q= Q_{\rm th}-Q_{\rm exp}$ and $\delta \tilde{Q}=\tilde{Q}_{\rm th}-Q_{\rm exp}$
are given in the last two columns.}

\begin{tabular}{|l |c |c |c |c |c |c|c|}
\noalign{\smallskip}\hline\noalign{\smallskip}
  nucl.  &$\lambda$  & $Q_{\rm exp}$& $Q_0$ &
$Q_{\rm th}$ &$\tilde{Q}_{\rm th}$& $\delta Q$ &$\delta \tilde{Q}$\\
\noalign{\smallskip}\hline\noalign{\smallskip}

$^{105}$In & $1g_{9/2}$& +0.83(5)&0.18 & +0.83 & 0.76 &0.00 &-0.07 \\

$^{107}$In & $1g_{9/2}$& +0.81(5) &0.18& +0.98 &  0.87&0.17 &  0.06   \\

$^{109}$In & $1g_{9/2}$& +0.84(3) &0.18& +1.11 &  1.00&0.27 &  0.16\\

$^{111}$In & $1g_{9/2}$& +0.80(2) &0.19 &+1.17 &  1.02&0.37&  0.22\\

$^{113}$In & $1g_{9/2}$& +0.80(4) &0.19& +1.12 &  0.98&0.32 &  0.16\\

$^{115}$In & $1g_{9/2}$& +0.81(5) &0.19 &+1.03 &  0.90&0.22 &  0.09 \\

&                       & 0.58(9)&   &&           &0.45 & 0.32\\

$^{117}$In & $1g_{9/2}$& +0.829(10)&0.19& +0.96   & 0.83 &0.131 &0.001  \\

$^{119}$In & $1g_{9/2}$& +0.854(7) &0.19 &+0.91   & 0.773&0.056 &-0.081\\

$^{121}$In & $1g_{9/2}$& +0.814(11) &0.19 &+0.833 & 0.711&0.019 &-0.103 \\

$^{123}$In & $1g_{9/2}$& +0.757(9)  &0.19 &+0.743 & 0.670&-0.014 &-0.087 \\

$^{125}$In & $1g_{9/2}$& +0.71(4)   &0.19 &+0.66  & 0.60 &-0.05 &-0.11 \\

$^{127}$In & $1g_{9/2}$& +0.59(3)   &0.19 &+0.55  & 0.52 &-0.04 &-0.07  \\

$^{115}$Sb & $2d_{5/2}$& -0.36(6)   &-0.14 &-0.88  & -0.62&-0.52 &-0.26  \\

$^{117}$Sb & $2d_{5/2}$&    -       &-0.14 &-0.817 & -0.585&- &-   \\

$^{119}$Sb & $2d_{5/2}$& -0.37(6)   &-0.14 &-0.76  & -0.55 &-0.39 &-0.18 \\

$^{121}$Sb & $2d_{5/2}$& -0.36(4)   &-0.14&-0.72  & -0.51 &-0.36 &-0.15 \\
&                      & -0.45(3)   & &      &       &-0.27 &-0.06 \\

$^{123}$Sb  & $1g_{7/2}$& -0.49(5)  &-0.17 &-0.74 & -0.55 &-0.25 &-0.06 \\

\hline
\end{tabular}
\label{tab:Q_p}
\end{table}

\begin{figure}[t]
  \centerline{\includegraphics[width=250pt]{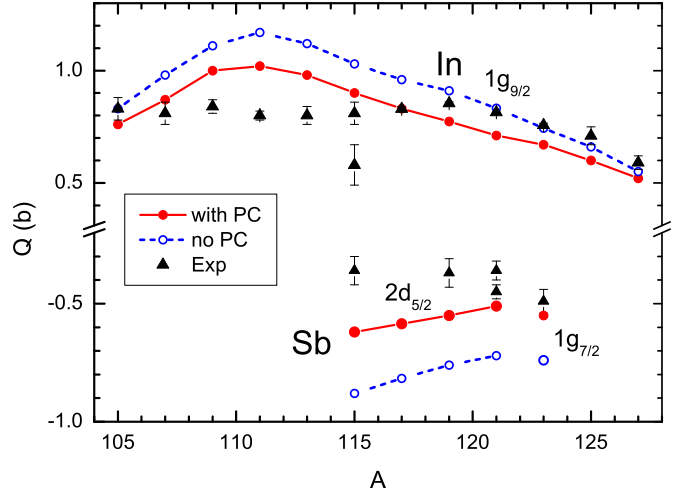}}
  \caption{Quadrupole moments of odd Sb and In isotopes with and without PC corrections. Experimental
  data are taken from \cite{Stone}.}
\end{figure}

\section{Charge radii of heavy Ca isotopes}

\begin{figure}[t]
  \centerline{\includegraphics[width=250pt]{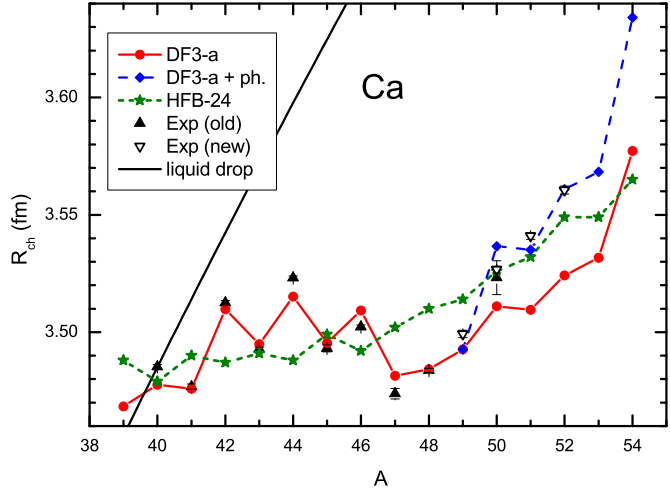}}
  \caption{ Charge radii of calcium isotopes.
Experimental data shown by closed (old) and open (new) triangles are taken from \cite{exp-Rch} and
\cite{exp-Ca-Rch}, respectively.}
\end{figure}

Recently, an anomalous $A$-dependence of the charge radii of calcium isotopes has been announced in
Ref. \cite{exp-Ca-Rch}. The results of the first high-precision measurements of the charge radii of
$^{49,51,52}$Ca nuclei were reported, and the article was titled ``Unexpectedly Large Charge Radii of
Neutron-Rich Calcium Isotopes.'' In \cite{Ca-Rch} this ``puzzle'' was resolved in terms of the PC
effects. A non-trivial point of this consideration is a special feature of the Fayans EDF which was
constructed in such a way that it should describe the data without PC corrections. Due to a fine
tuning of the parameters of the surface term of the EDF DF3, the authors of \cite{Fay} managed to
reproduce with high accuracy very fancy $A$-dependence of the calcium charge radii from $^{40}$Ca till
$^{48}$Ca at the mean-field level. The corresponding values of $R_{\rm ch}$ in \cite{Fay} coincide
practically with DF3-a those in Fig. 6. As far as the charge radius of the nucleus $^{48}$Ca is
reproduced with DF3-a EDF without phonons, such an ansatz was suggested in \cite{Ca-Rch} for isotopes
with $A{>}48$: \beq \tilde{\delta}^{\rm PC} \langle r_{\rm ch}^2 \rangle(^A{\rm Ca})  \equiv
\delta^{\rm PC} \langle r_{\rm ch}^2 \rangle(^A{\rm Ca}) - \delta^{\rm PC} \langle r_{\rm ch}^2
\rangle(^{48}{\rm Ca}), \label{dRA48}\eeq with obvious notation. This ansatz is a direct application
of the idea to separate the fluctuated part of the PC correction. Values of $R_{\rm ch}(A{>}48)$ with
the PC corrections are found with the use of Eq. (\ref{dRA48}). The $3^-_1$ and $2^+_1$ phonons in Ca
isotopes are considered. All of them are surface vibrations with well defined surface peaks similar
that in Fig. 1. If one neglects the in-volume term $\chi_L$ in Eq. (\ref{gL}), the BM model formula
\cite{BM2} can be obtained: \beq \delta \langle r^2\rangle_L = R_0^2 \frac 5 {4\pi} \beta_L^2,
\label{dR_BM} \eeq where $R_0=1.2 A^{1/3}\;$fm, and $\beta_L$ is the parameter of the ``dynamical
deformation'' related to the coefficient $\beta_L$ in (\ref{gL}) as follows: $\alpha_L=\beta_L R_0 /
\sqrt{2L+1}$. To find the total PC correction $\delta^{\rm PC} \langle r_{\rm ch}^2\rangle$ for a
nucleus under consideration, one should to sum the values of (\ref{dR_BM}) for $L{=}2$ and for
$L{=}3$. Results of calculations on the base of Eqs. (\ref{dRA48}), (\ref{dR_BM}) for the Fayans EDF
DF3-a are presented in Fig. 6. For a comparison, predictions are shown of the Skyrme EDF HFB-24
\cite{HFB}, belonging to the family of Skyrme EDFs HFB-($17\div27$), which are champions in accuracy
of self-consistent description of nuclear masses. We see that, at the mean-field level, description of
the charge radii of Ca isotopes with the DF3-a EDF is much better than that with the HFB-24 EDF.
However, for the isotopes $^{50-52}$Ca the experimental points of \cite{exp-Ca-Rch}, indeed, there are
noticeably higher than those of the DF3-a EDF. We see that the account for the fluctuating part of the
PC corrections with the method described above does solve the problem.

\section{Conclusions}

  A brief review is presented of recent results of describing the PC effects in odd magic and semi-magic
nuclei within the self-consistent TFFS. The perturbation theory in $g_L^2$ is used, $g_L$ being the
vertex of creating the $L$-phonon. In addition to the usual pole diagrams, the non-pole ones, also
proportional to $g_L^2$, are considered. Their contributions are often of a crucial importance. PC
corrections to the single-particle energies for $^{40}$Ca and $^{208}$Pb found in \cite{Levels} are
presented based on the Fayans EDF DF3-a \cite{DF3-a}. For the lead nucleus, with 24 experimental
 SP energies $\eps_{\lambda}$ known, the average difference between theoretical and experimental
 values is $\langle\delta \eps_{\lambda}\rangle_{\rm rms}{=}0.51\;$MeV without PC corrections
and 0.34 MeV with PC corrections. For a comparison, the corresponding value for the popular Skyrme EDF
HFB-17 EDF is equal to 1.15 MeV.

The results of the first self-consistent description of quadrupole moments of odd semi-magic nuclei
with accounting for the PC corrections in \cite{PC-Q} are also presented. The odd In and Sb isotopes,
the odd-proton neighbors of even Sn isotopes, are considered. Two main PC corrections in this case,
the ``end correction'' and the one of the induced interaction, are of opposite signs and cancel each
other strongly. Therefore, the complete self-consistency of the calculation scheme is of primary
importance. Another peculiarity of this problem is related to the PC correction due to the quadrupole
moment of the $L$-phonon. The usual triangle diagram for this process, the left part of Fig. 4,
contains a term with a singular behavior at small phonon excitation energy $\omega_L$. The same
singularity, with opposite sign, persists in the non-pole analogue of this diagram, the right part of
Fig. 4. Their sum is already regular at small $\omega_L$. Account for the PC corrections makes the
overall agreement with the data significantly better. For 16 nuclei considered, the rms value of the
difference between the theoretical and experimental values of the quadrupole moment $Q$ is
$\langle\delta \tilde{Q}\rangle_{\rm rms}=0.15\;$b with PC corrections and  0.27 b, without.

At last, some results are presented of Ref. \cite{Ca-Rch}, where recently announced problem of
extremely high values charge radii of heavy Ca isotopes \cite{exp-Ca-Rch} is solved in terms of a
consistent consideration of the PC effects.

\section{Acknowledgments}
The work is supported by the Russian Science Foundation, Grants Nos. 16-12-10155 and 16-12-10161. It
was also partially supported by the RFBR Grant 16-02-00228. Calculations were partially carried out at
the Computer Center of NRC 'Kurchatov Institute'.

% References

%\nocite{*}
%\bibliographystyle{aipnum-cp}%
%\bibliography{Saperstein}%

\end{document}